# UO$_2$/BeO interfacial thermal resistance and its effect on fuel thermal conductivity

Xueyan Zhu [a,b], Rui Gao [c], Hengfeng Gong [d], Tong Liu [d], De-Ye Lin [a,b,*] and Haifeng Song [a,b,*]

[a] CAEP Software Center for High Performance Numerical Simulation, Beijing 100088, P.R. China

[b] Institute of Applied Physics and Computational Mathematics, Beijing 100088, P.R. China

[c] Institute of Materials, China Academy of Engineering Physics, Jiangyou, Sichuan 621908, P.R. China

[d] China Nuclear Power Technology Research Institute, Shenzhen 518031, P.R. China

**Abstract**

UO$_2$/BeO interfacial thermal resistance (ITR) is calculated by diffuse mismatch model (DMM) and the effects of ITR on UO$_2$-BeO thermal conductivity are investigated. ITR predicted by DMM is on the order of $10^{-9}$ m$^2$K/W. Using this ITR, UO$_2$-BeO thermal conductivities are calculated by theoretical models and compared with experimental data. The results of this comparison indicate that DMM prediction is applicable to the interface between UO$_2$ and dispersed BeO, while not applicable to the interface between UO$_2$ and continuous BeO. If the thermal conductivity of UO$_2$ containing continuous BeO was to be in agreement with experimental data, its ITR should be on the order of $10^{-6} - 10^{-5}$ m$^2$K/W. Therefore, the vibrational mismatch between UO$_2$ and BeO considered by DMM is the major mechanism for attenuating the heat flux through UO$_2$/dispersed-BeO interface, but not for UO$_2$/continuous-BeO interface. Furthermore, it is found that the presence of ITR leads to the dependence of the thermal conductivity of UO$_2$ containing dispersed BeO on BeO size. With the decrease in BeO size, UO$_2$-BeO thermal conductivity decreases. When BeO size is smaller than a critical value, UO$_2$-BeO thermal conductivity becomes even smaller than UO$_2$ thermal conductivity. For UO$_2$ containing continuous BeO, the thermal conductivity decreases with the decrease in the size of UO$_2$ granule surrounded by BeO, but not necessarily smaller than UO$_2$ thermal conductivity. Under a critical temperature, UO$_2$-BeO thermal conductivity is always larger than UO$_2$ thermal conductivity. Above the critical temperature, UO$_2$-BeO thermal conductivity is larger than UO$_2$ thermal conductivity only when UO$_2$ granule size is large enough. The conditions for achieving the targeted enhancement of UO$_2$ thermal conductivity by doping with BeO are derived. These conditions can be used to design and optimize the distribution, content, size of BeO, and the size of UO$_2$ granule.

* lin_deye@iapcm.ac.cn
* song_haifeng@iapcm.ac.cn

## 1. Introduction

Uranium dioxide ($UO_2$) is the most widely used fuel in commercial nuclear plants. However, its low thermal conductivity has led to a variety of problems related to the performance and safety of the reactor, such as large centerline temperatures, pellet cracking and fuel relocation. An approach that has been demonstrated to be effective in improving the thermal conductivity of $UO_2$ is to incorporate a high conductivity phase in $UO_2$ fuel [1-8]. Beryllium oxide (BeO) is a promising additives due to its high thermal conductivity, low thermal neutron absorption cross-section, good chemical compatibility with $UO_2$, and high resistance to water steam. Significant efforts have been made to investigate the effects of BeO content and distribution on $UO_2$ thermal conductivity [1-6]. It was demonstrated that the increase in BeO content can lead to an effective improvement of $UO_2$ thermal conductivity, and a continuous BeO phase leads to a higher thermal conductivity than dispersed BeO. In addition to BeO content and distribution, the quality of $UO_2$/BeO interface is also crucial that affects the thermal conductivity of this composite [3, 6]. However, this has been little concerned by previous work. Interfacial thermal resistance (ITR) is the parameter that characterizes an interface's resistance to heat flow. The work of this manuscript focuses on the calculation of $UO_2$/BeO ITR through theoretical models and the investigation of the effect of ITR on $UO_2$-BeO thermal conductivity.

The importance of ITR effect on $UO_2$-BeO thermal conductivity has been indicated by several studies. Latta *et al.* [3] found the overestimation of $UO_2$-BeO thermal conductivity by 2D finite element modelling (FEM) without considering ITR compared with the measured thermal conductivity of a green granule $UO_2$-BeO sample. They attributed this difference to the ignorance of $UO_2$/BeO ITR in FEM. Badry *et al.* [6] also reported the overestimation of the thermal conductivities of both $UO_2$ containing dispersed and continuous BeO by FEM without considering ITR.

Although thermal conductivity models that can consider ITR effect have been used for predicting the thermal conductivities of composite fuel, the value of ITR has not been determined reasonably. Yeo [9] and Liu *et al.* [10] used the model derived by Hasselman and Johnson [11] to calculate the thermal conductivities of $UO_2$-SiC and $UO_2$-BeO, respectively.

However, the ITR was determined by acoustic mismatch model (AMM) [12] under Debye approximation in the low temperature limit, which cannot be applied to the temperature range concerned in the reactor. Badry *et al.* [6] obtained $UO_2$/BeO ITR through fitting the experimental value of $UO_2$-BeO thermal conductivity by their proposed FEM model. Nevertheless, the model is 2D and assumes a uniform distribution of dispersed BeO, which can lead to an underestimation of the thermal conductivity. And only one data point for 5 vol.% dispersed microstructure at 200 °C was used for the fit.

Two common theoretical models for determining ITR are AMM [12] and diffuse mismatch model (DMM) [13]. AMM assumes a perfect interface, and treats the phonons as plane waves and the materials as continua. Therefore, the phonons specularly reflect or transmit through the interface in AMM. Later studies found that AMM breaks down for phonons with high frequency, which can be produced at high temperatures and scattered diffusely by a rough interface. To account for this type of scattering, DMM was developed, which assumes that ITR is caused by the vibrational mismatch between the two materials and a phonon incident on the interface forgets where it came from. As a result, the phonon transmission probability is determined by the ratio of the phonon density of states of the materials on both sides of the interface. These comparisons between AMM and DMM indicate that DMM is more suitable for predicting the ITR in the reactor, where the temperature is high and the interface is rough. AMM and DMM were originally proposed based on a linear dispersion relationship (Debye approximation) for simplicity. Recently, AMM and DMM based on full-band (FB) phonon dispersions have also been developed [14, 15].

In this work, $UO_2$/BeO ITR is predicted by DMM based on full-band phonon dispersions that are calculated by density functional theory (DFT). To examine if DMM grasps the major mechanism for attenuating the heat flux through practical $UO_2$/BeO interface, $UO_2$-BeO thermal conductivities calculated by theoretical models considering the ITR are compared with experimental data. Then, the impacts of ITR on $UO_2$-BeO thermal conductivity are investigated. And the conditions for the targeted enhancement of $UO_2$ thermal conductivity by doping with dispersed and continuous BeO are derived, respectively.

This manuscript is organized as follows. In Section 2, details of computational models and methods are presented. In Section 3.1, $UO_2$/BeO ITRs calculated by DMM based on Debye approximation, acoustic modes and FB dispersions are compared and discussed. Then, comparisons are made between the thermal conductivities predicted by theoretical models considering the ITR and experimental data for $UO_2$ containing dispersed BeO and continuous BeO in Section 3.2 and Section 3.3, respectively. Meanwhile, the impacts of ITR on $UO_2$-BeO thermal conductivity are discussed. Finally, conclusions are presented in Section 4.

## 2. Model and Methods

Phonon dispersions of $UO_2$ and wurtzite-BeO are obtained through lattice dynamics calculations at harmonic level using the Phonopy code [16]. The second-order interatomic force constants (IFCs) required for phonon calculations are obtained through DFT calculations.

DFT calculations are carried out through the Vienna Ab-initio Simulation Package (VASP) [17, 18]. Electron-ion interaction is described by the projector augmented wave (PAW) method [19, 20]. The generalized gradient approximation (GGA) as parameterized by Perdew, Burke and Ernzerhof (PBE) [21] is used to calculate the electron-electron exchange-correlation energy. For $UO_2$, the strong on-site Coulomb repulsion of 5f electron is considered through Hubbard U method [22] with $U = 4.5$ eV and $J = 0.51$ eV. The energy cutoff of plane wave basis set is set to 500 eV and 520 eV for $UO_2$ and BeO, respectively. For performing the integration over eigenvalues, the Gaussian function with a smearing width of 0.05 eV for $UO_2$ and tetrahedron method for BeO are used. Based on the relaxed unitcell, 2×2×2 supercell of $UO_2$ and 4×4×4 supercell of BeO are created for calculating the second-order IFCs. The Brillouin zone is sampled with a gamma-centered k-point mesh of 4×4×4 for $UO_2$ and 3×3×2 for BeO, respectively. The criteria for the convergence of electronic self-consistent iterations is set to 0.01 meV. The phonon dispersion calculations are performed on a q-point mesh of 20×20×20 with the polarization effects included. The Born effective charges are 5.50 e for $U^{4+}$ and -2.75 e for $O^{2-}$ in $UO_2$, and (1.78 e, 1.85 e) for $Be^{2+}$ and (-1.78 e, -1.85 e) for $O^{2-}$ in BeO.

With the full-band dispersion of $UO_2$ and BeO, the energy transmission probability from $UO_2$ to BeO is calculated by the following formula according to the approximation of DMM [13, 14]:

$$\alpha_{UO2\to BeO}^{DMM}(\omega') = \frac{\int_{BeO} \sum_j |\mathbf{v}(\mathbf{q},i)\cdot\mathbf{n}|\delta(\omega'-\omega(\mathbf{q},i))d\mathbf{q}}{\int_{UO2} \sum_i |\mathbf{v}(\mathbf{q},i)\cdot\mathbf{n}|\delta(\omega'-\omega(\mathbf{q},i))d\mathbf{q} + \int_{BeO} \sum_j |\mathbf{v}(\mathbf{q},i)\cdot\mathbf{n}|\delta(\omega'-\omega(\mathbf{q},i))d\mathbf{q}}, \quad (1)$$

where, $\mathbf{v}(\mathbf{q}, i)$ and $\omega(\mathbf{q}, i)$ are the group velocity and frequency of phonon with wave vector $\mathbf{q}$ and polarization $i$, respectively. $\mathbf{n}$ is a unit vector normal to the $UO_2$/BeO interface. Then, the thermal boundary conductance of $UO_2$/BeO interface is obtained by:

$$G_{UO2\to BeO} = \frac{1}{2}\frac{1}{8\pi^3}\sum_i \int_\mathbf{q} \frac{1}{k_B T^2} \alpha_{UO2\to BeO}^{DMM}(\mathbf{q},i)$$
$$\times (\hbar\omega(\mathbf{q},i))^2 |\mathbf{v}(\mathbf{q},i)\cdot\mathbf{n}| \frac{\exp\left(\frac{\hbar\omega(\mathbf{q},i)}{k_B T}\right)}{\left[\exp\left(\frac{\hbar\omega(\mathbf{q},i)}{k_B T}\right)-1\right]^2} d\mathbf{q}, \quad (2)$$

where, $k_B$ is the Boltzmann constant, $\hbar$ is the reduced Planck constant, and $T$ is the temperature. The inverse of the thermal boundary conductance is the interfacial thermal resistance (ITR).

For simplicity, early studies on thermal boundary conductance were usually based on a linear dispersion relationship (Debye approximation). Under Debye approximation, the transmission probability of DMM is calculated by [13]:

$$\alpha_{UO2\to BeO}^{DMM/Debye} = \frac{\sum_i c_{BeO,i}^{-2}}{\sum_i c_{BeO,i}^{-2} + \sum_i c_{UO2,i}^{-2}}, \quad (3)$$

where, $c$ is the sound velocity. The corresponding thermal boundary conductance is

$$G_{UO2\to BeO}^{Debye} = \frac{1}{4\pi^2}\frac{k_B^4}{\hbar^3}T^3 \sum_i c_i^{-2}\Gamma_i \int_0^{\frac{\hbar\omega_D}{k_B T}} \xi^4 \frac{\exp(\xi)}{[\exp(\xi)-1]^2}d\xi,$$
$$\Gamma_i = \int_0^{\pi/2} \alpha_{UO2\to BeO}^{Debye} \sin\theta\cos\theta d\theta \quad (4)$$

in which, $\omega_D$ is the Debye frequency.

## 3. Results and discussion

*3.1. Interfacial thermal resistance of UO₂/BeO calculated by DMM*

The phonon dispersions of $UO_2$ and BeO calculated by DFT are plotted in Fig. 1. Results of this work (black lines) are in good agreement with experimental data [23, 24] (red points). As shown in Fig. 2, the density of states of $UO_2$ and BeO mainly overlap in the region around 10 – 17 THz. The highest frequency of acoustic modes in $UO_2$ and BeO is 5.914 THz and

17.143 THz, respectively. This indicates that heat transfers from $UO_2$ to BeO mainly through transforming the optic modes of $UO_2$ to the acoustic modes of BeO on the interface according to the approximation of DMM.

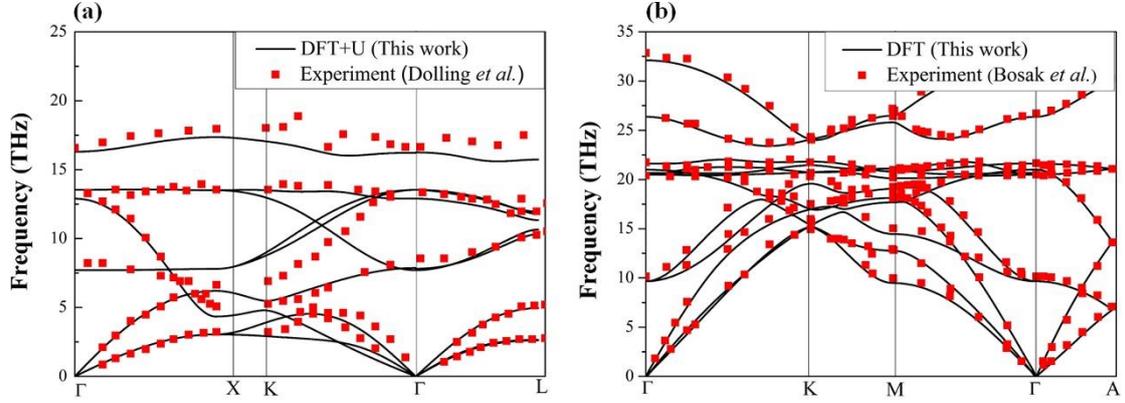

**Fig. 1.** Phonon dispersions of $UO_2$ (a) and BeO (b) along high symmetric directions. The solid lines show the prediction of this work. The red points represent experimental data [23, 24].

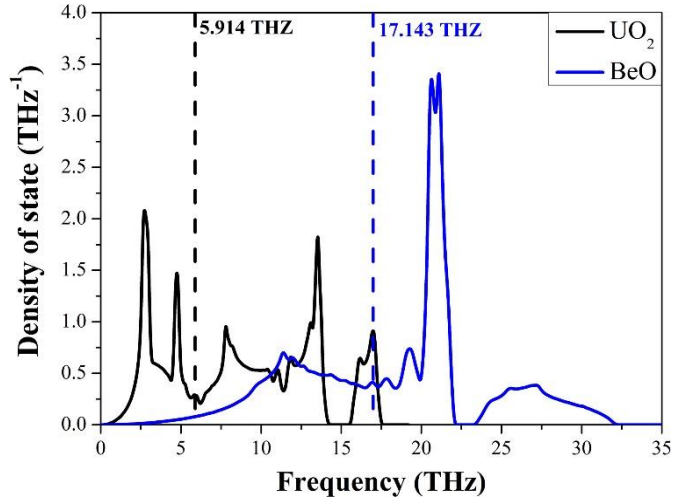

**Fig. 2.** Phonon density of states of $UO_2$ and BeO.

ITRs of $UO_2(111)/BeO(0001)$ interface are calculated using DMM based on FB phonon dispersions, acoustic modes, and Debye approximation, respectively, as shown in Fig. 3(a). At low temperatures ($T < 45$ K), the difference between ITRs calculated through DMM using only the acoustic phonon modes and full bands of phonons is within 20%. This is because most phonons have not been activated and stay in the acoustic modes with low energy at low temperatures. The results of DMM based on Debye approximation are higher than those based on FB dispersions and acoustic modes. This difference is due to the fact that Debye approximation fails to model the lowering of group velocities away from the Brillouin zone center. When the temperature is high ($T > 70$ K), ITRs calculated by DMM based on acoustic modes and Debye approximation become one order of magnitude larger than that based on full bands. This is because the phonons of optic modes in $UO_2$ are activated at high temperatures and become the main carriers of heat through the interface, which is indicated by the overlay of the density of states of $UO_2$ and BeO that has been discussed above. These comparisons demonstrate that using full bands of phonon dispersions is necessary for calculating the ITR of $UO_2$/BeO interface in the reactor, where the temperature is usually larger than 600 K.

ITRs of $UO_2$/BeO interface with different crystal orientations are investigated using DMM based on full bands of phonon dispersions as shown in Fig. 3(b). All the ITRs decrease with the increase in temperature and on the order of $10^{-9}$ m$^2$K/W. The orientation of the interface has a small impact on ITR. The largest ITR is obtained in the case of $UO_2(111)$/BeO(0001) interface, while the lowest ITR presents in the case of $UO_2(111)$/BeO(11$\bar{2}$0) interface. ITRs of $UO_2(111)$/BeO(10$\bar{1}$0) and $UO_2(111)$/BeO(10$\bar{1}$1) interfaces are in close agreement with each other, and between the ITRs of $UO_2(111)$/BeO(0001) and $UO_2(111)$/BeO(11$\bar{2}$0) interfaces.

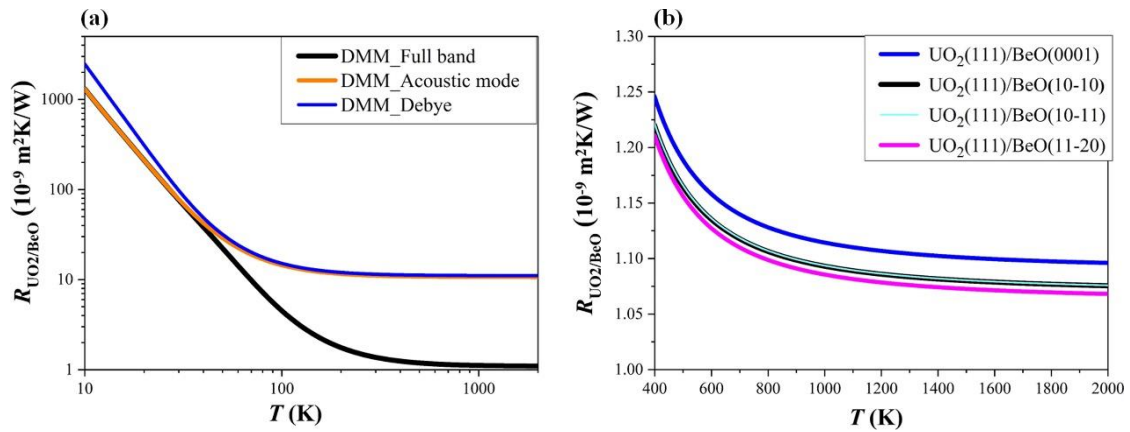

**Fig. 3.** Interfacial thermal resistance (ITR) of $UO_2$/BeO interface. (a) Comparison of the ITRs calculated using DMM based on FB phonon dispersions, acoustic modes and Debye approximation, respectively. (b) Comparison of the ITRs for $UO_2$/BeO interface with different crystal orientations.

*3.2. Thermal conductivity of $UO_2$ containing dispersed BeO*

Although DMM can consider diffuse scattering caused by a rough interface, the ITR of practical $UO_2$/BeO interface may be influenced by many other factors, such as, inelastic scattering, surface roughness, void existing on the interface, etc. Therefore, it is necessary to examine if the ITR predicted by DMM is applicable to real $UO_2$/BeO interface. In this and next section, thermal conductivities calculated by theoretical models using the ITR predicted by DMM are compared with experimental data for $UO_2$ doped with dispersed and continuous BeO, respectively. Then, the effects of ITR on the thermal conductivity of these two types of $UO_2$-BeO are investigated and discussed, respectively.

The thermal conductivity model for a matrix with randomly distributed spherical dispersions that can consider ITR effect has been derived by Hasselman and Johnson (H-J model) [11]:

$$k_{eff} = k_m \frac{2 \cdot \left(\frac{k_p}{k_m} - \frac{k_p}{a} \cdot R - 1\right) \cdot V_p + \frac{k_p}{k_m} + 2 \cdot \frac{k_p}{a} \cdot R + 2}{\left(1 - \frac{k_p}{k_m} + \frac{k_p}{a} \cdot R\right) \cdot V_p + \frac{k_p}{k_m} + 2 \cdot \frac{k_p}{a} \cdot R + 2} \qquad (5)$$

where, $k_m$ and $k_p$ are the thermal conductivities of the matrix and the spherical dispersions, respectively, $V_p$ is the volume fraction of the dispersions, $a$ is the dispersion radius, and $R$ is the ITR of the interface between the matrix and the dispersion. For $UO_2$ containing dispersed BeO, $UO_2$ is the matrix, the thermal conductivity of which is calculated using the model recommended by Fink [25]. The thermal conductivity of dispersed BeO is determined by the experimental results of Takahashi and Murabayashi [26].

First, comparison is made between $UO_2$-BeO thermal conductivity calculated using the above model with ITR setting to zero and experimental data. $UO_2$-BeO with BeO content of 4.2 vol.%, 5 vol.% and 10 vol.% are investigated to compare with the data from different experiments. All the experimental data are corrected to 100% theoretical density using the equation recommended by Brandt and Neuer [27]. As shown in Fig. 4(a), the results of H-J model with $R_{UO2/BeO} = 0$ are in good agreement with the experimental data from Ishimoto *et al.* [1]. However, the difference between the predictions of H-J model with $R_{UO2/BeO} = 0$ and experimental results of Badry *et al.* [6] is notable (Fig. 4(b)). The relative difference varies from 5% to 18% for $UO_2$-5vol.%BeO and 5% to 14% for $UO_2$-10vol.%BeO. Without considering ITR, H-J model overestimates the thermal conductivities at all temperatures.

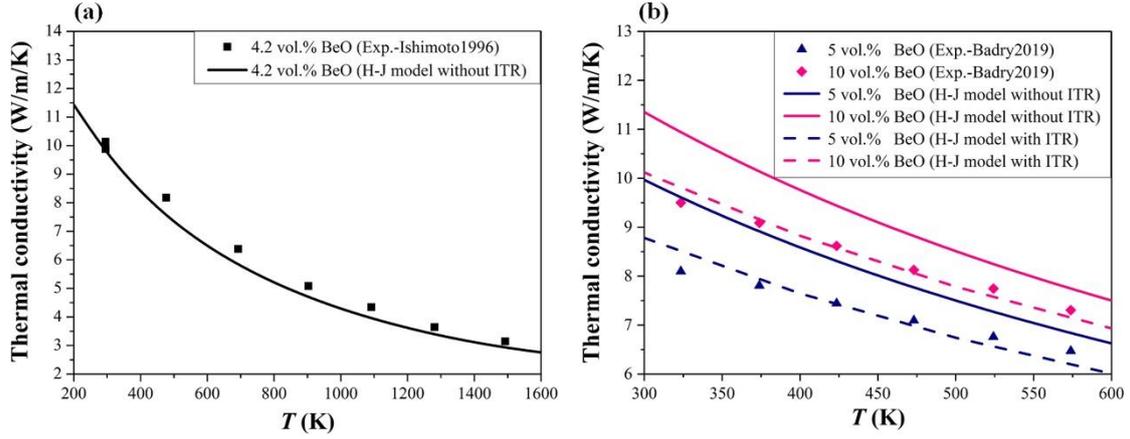

**Fig. 4.** Comparisons between the thermal conductivities of $UO_2$ doped with dispersed BeO calculated by H-J model (lines) and several experimental data [1, 6] (filled points). The solid lines are model predictions without considering ITR, and the dotted lines are model predictions considering ITR.

The above comparisons indicate that the ITR effect on $UO_2$-BeO thermal conductivity is different for the $UO_2$-BeO samples in different experiments. It is noticed from Eq. (5) that the effect of ITR is closely related to the radius of the dispersion $a$. Only when $a/R < k_p$, ITR would have a significant effect on the thermal conductivity. For $UO_2$-BeO, ITR calculated by full-band DMM is on the order of $10^{-9}$ m$^2$K/W. $k_p = k_{BeO} \sim 10^2$ W/m/K. Therefore, the ITR impact should be noticeable when $a < 100$ nm. BeO radius in the sample of Ishimoto *et al.* [1] is on the order of μm, which leads to the negligible effect of ITR, thus, explaining the good agreement between Ishimoto's experimental data and the results of H-J model with ITR setting to zero. The noticeable difference between Badry's experimental data [6] and predictions of H-J model without ITR indicates that BeO radius in the sample of Badry *et al.* should be smaller than 100 nm.

Badry's experimental data are fitted by H-J model with ITR setting to DMM predictions ($R = 10^{-9}$ m$^2$K/W) as shown by the dotted lines in Fig. 4(b). The fitted BeO radius is obtained to be 11.99 nm and 44.12 nm for 5 vol.% BeO and 10 vol.% BeO, respectively, which is on the same order of magnitude with BeO radius (12.6 nm and 17.84 nm) reported by Badry *et al.*. This indicates that the ITR predicted by DMM is applicable to the real interface between $UO_2$ and dispersed BeO. Therefore, the heat flux through $UO_2$/dispersed-BeO interface is mainly attenuated by the vibrational mismatch between $UO_2$ and BeO.

ITR effect on UO$_2$-BeO thermal conductivity is further investigated. H-J model expressed by Eq. (5) indicates that the presence of ITR leads to the dependence of UO$_2$-BeO thermal conductivity on BeO radius. Using H-J model with $R = 0.1\times10^{-9}$, $1.0\times10^{-9}$, and $3.0\times10^{-9}$ m$^2$K/W, thermal conductivities of UO$_2$ with BeO content of 5 vol.% at 323.15 K and 573.15 K are calculated. As shown in Fig. 5, with the decrease in BeO radius, UO$_2$-BeO thermal conductivity decreases. When BeO radius is smaller than a critical value, UO$_2$-BeO thermal conductivity becomes even smaller than UO$_2$. The equation for this critical radius can be derived by equating UO$_2$-BeO thermal conductivity $k_{eff}$ (Eq. (5)) to UO$_2$ thermal conductivity $k_m$:

$$a_{\text{critical}} = \frac{k_p \cdot R}{k_p / k_m - 1}. \tag{6}$$

Therefore, due to the existence of ITR, the thermal conductivity of UO$_2$ can only be enhanced by BeO dispersions with radius satisfying the following condition:

$$a > a_{\text{critical}}. \tag{7}$$

Setting $R$ to DMM predictions (10$^{-9}$ m$^2$K/W), the variation of $a_{\text{critical}}$ with temperature is plotted in Fig. 6 (black points). With the increase in temperature, $a_{\text{critical}}$ decreases. As shown by the red line in Fig. 6, $a_{\text{critical}}$ vs. $T$ can be fitted by the empirical function 1/(a+b$T$) with the fitted parameters a = $5.93\times10^{-2}$ nm$^{-1}$ and b = $1.78\times10^{-4}$ K$^{-1}$nm$^{-1}$. This empirical function can be conveniently used for estimating $a_{\text{critical}}$.

Furthermore, the condition for the targeted improvement of UO$_2$ thermal conductivity by doping with dispersed BeO are derived. If UO$_2$ thermal conductivity was to be increased by at least $\xi$, that is $k_{eff} \geq (1+\xi)k_m$, the following conditions should be satisfied according to Eq. (5):

$$\begin{aligned} V_p &\geq \frac{\xi}{3+\xi} \frac{k_p/k_m + 2}{k_p/k_m - 1} \\ a &\geq \frac{k_p R}{f} \left[ 2\xi + (3+\xi)V_p \right] \\ f &= V_p(3+\xi)(k_p/k_m - 1) - \xi(k_p/k_m + 2) \end{aligned} \tag{8}$$

For example, if UO$_2$ thermal conductivity was to be improved by at least 40% at 1000 K, the volume fraction of doped BeO needed to be larger than 14.47%. If in the real experiments, 15 vol.% BeO was inserted, the radius of BeO needed to be larger than 304.31 nm when $R = 10^{-9}$ m$^2$K/W for achieving 40% enhancement of the thermal conductivity.

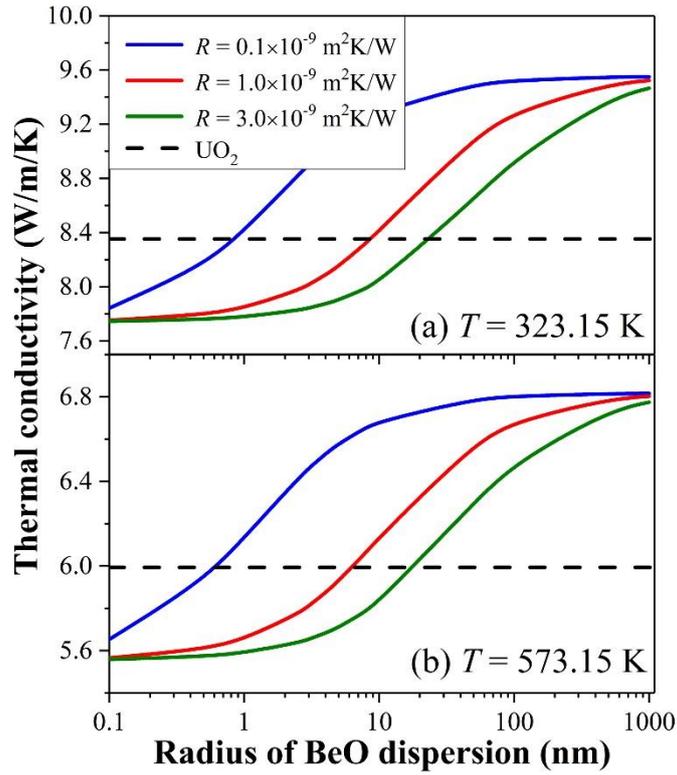

**Fig. 5.** Variation of the thermal conductivity of UO$_2$ containing 5 vol.% dispersed BeO with respect to BeO radius.

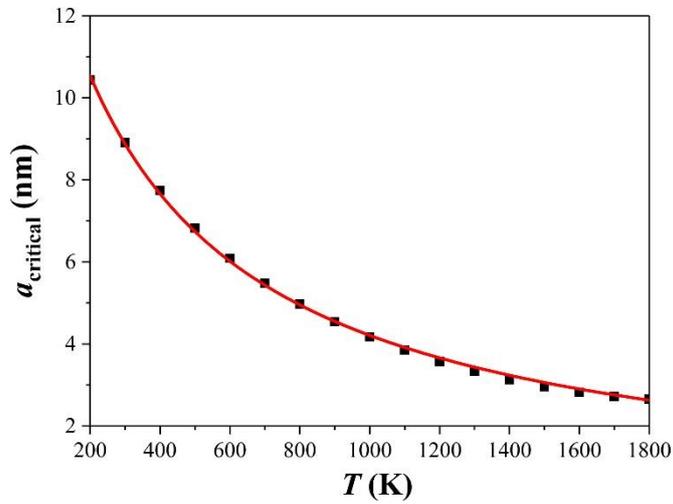

**Fig. 6.** Variation of the critical radius of BeO with temperature. The black points are calculated by Eq. (6), which are fitted by the function $1/(a+bT)$ as shown by the red line.

*3.3 Thermal conductivity of UO$_2$ containing continuous BeO*

UO$_2$ containing continuous BeO is modeled by 2D and 3D grid pattern geometries as shown in Fig. 7. Using the method of equivalent thermal resistance, the corresponding 2D and 3D thermal conductivity models are derived:

2D model

$$k_{eff} = \frac{1}{l_m/l_p + 1} k_p + \frac{1}{k_p/k_m + l_p/l_m + 2k_pR/l_m} k_p, \quad (9)$$

$$\frac{l_m}{l_p} = \frac{(1-V_p)^{1/2}}{1-(1-V_p)^{1/2}}$$

3D model

$$k_{eff} = \left(\frac{1}{l_m/l_p + l_p/l_m + 2} + \frac{1}{l_m/l_p + 1}\right)k_p + \frac{1}{k_p/k_m + l_p/l_m + 2k_pR/l_m} \frac{1}{1 + l_p/l_m} k_p, \quad (10)$$

$$\frac{l_m}{l_p} = \frac{(1-V_p)^{1/3}}{1-(1-V_p)^{1/3}}$$

where, $k_m$ and $k_p$ are the thermal conductivities of UO$_2$ matrix and BeO second phase, respectively, $V_p$ is the volume fraction of BeO, $l_m$ is the granule size of UO$_2$ matrix surrounded by BeO, and $R$ is UO$_2$/BeO ITR.

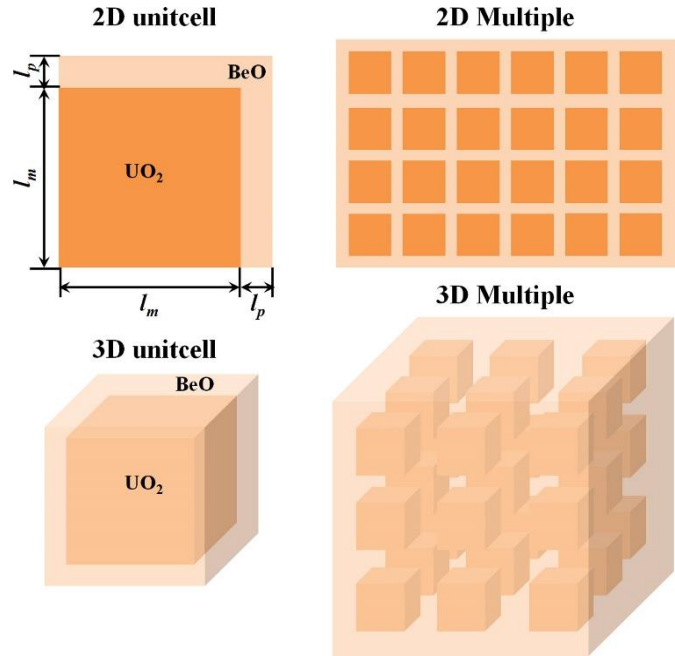

**Fig. 7.** 2D and 3D models for the theoretical calculation of the thermal conductivity of UO$_2$ containing continuous BeO.

To verify the above models, model predictions are compared with the results of 2D FEM reported by Latta *et al.* [3]. As shown in Fig. 8, the predictions of 2D model are in good agreement with the results of 2D FEM. The predictions of 3D model show larger values, which is reasonable.

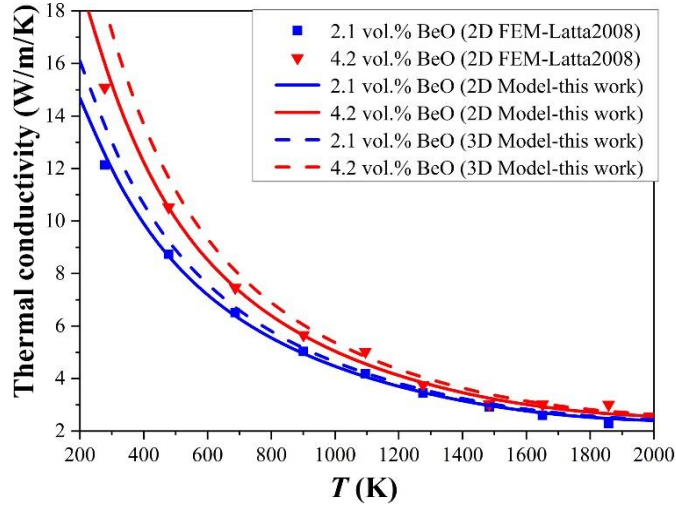

**Fig. 8.** Comparison between the thermal conductivity of $UO_2$ containing continuous BeO predicted by Eq. (9) and Eq. (10) and the results of 2D FEM [3].

The thermal conductivities calculated by the above 3D model with ITR setting to zero are compared with experimental data [1, 4]. As shown in Fig. 9(a), the thermal conductivities calculated by the 3D model without considering ITR are all larger than the experimental data [1, 4]. This indicates that ITR is a non-negligible factor that influences the thermal conductivity of $UO_2$ containing continuous BeO.

Setting ITR to DMM predictions ($R = 10^{-9}$ $m^2K/W$), $UO_2$-BeO thermal conductivities are further calculated by 3D model. The size of $UO_2$ granule surrounded by continuous BeO is set to 100 μm according to the experiments. It is found that there is negligible decrease of $UO_2$-BeO thermal conductivity compared with that calculated without considering ITR. This indicates that the ITR predicted by DMM is not applicable to the real interface between $UO_2$ and continuous BeO. If $UO_2$-BeO thermal conductivity was to be in agreement with experimental data (Fig. 9(b)), $UO_2$/BeO ITR should be on the order of $10^{-6}$ ~ $10^{-5}$ $m^2K/W$. Therefore, the vibrational mismatch between $UO_2$ and BeO should not be the major mechanism for attenuating the heat flux through $UO_2$/continuous-BeO. There may exist voids on the interface that largely attenuate the heat flux. This result indicates that the thermal resistance of the interface between $UO_2$ and continuous BeO may be contact resistance.

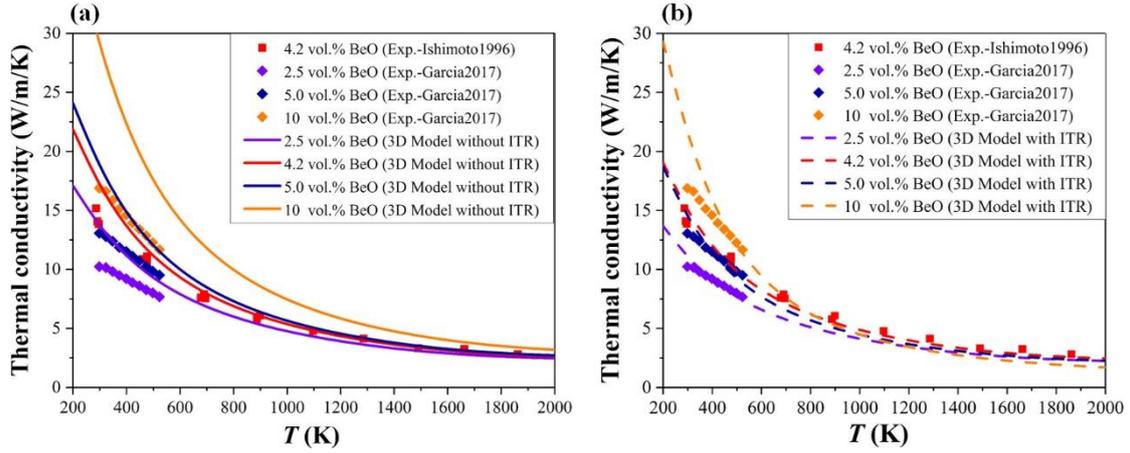

**Fig. 9.** Comparisons between the thermal conductivities of UO$_2$ doped with continuous BeO calculated by the 3D model expressed by Eq. (10) (lines) and several experimental data [1, 4] (filled points). The solid lines are model predictions without considering ITR, and the dotted lines are model predictions considering ITR.

From the models expressed by Eqs. (9-10), it is known that the presence of ITR leads to the dependence of UO$_2$-BeO thermal conductivity on UO$_2$ granule size $l_m$. Using the 3D model with $R = 10^{-6}$ and $10^{-5}$ m$^2$K/W, thermal conductivities of UO$_2$-BeO with BeO content of 5 vol.% at 323.15 K and 573.15 K are calculated, respectively. As shown in Fig. 10, the thermal conductivity decreases with the decrease in $l_m$. Interestingly, UO$_2$-BeO thermal conductivity at 323.15 K is larger than UO$_2$ thermal conductivity no matter how small the size of UO$_2$ granule is. However, at 573.15 K, UO$_2$-BeO thermal conductivity becomes smaller than UO$_2$ thermal conductivity when $l_m$ is smaller than a critical value, which is 4.06 μm and 40.55 μm for $R = 10^{-6}$ and $10^{-5}$ m$^2$K/W, respectively. The equation for this critical granule size can be derived by equating UO$_2$-BeO thermal conductivity $k_{eff}$ (Eq. (10)) to UO$_2$ thermal conductivity $k_m$:

$$l_{m\_critical} = \frac{Rl_{mp}\left[-2k_m\left(1+l_{mp}\right)^2 + k_p\left(2+4l_{mp}\right)\right]}{\left(l_{mp}^2 - l_{mp} - 1\right) + k_m\left(l_{mp}+1\right)^2/k_p - k_p\left(2l_{mp}^2 + l_{mp}\right)/k_m}$$

$$l_{mp} = \frac{l_m}{l_p} = \frac{\left(1-V_p\right)^{1/3}}{1-\left(1-V_p\right)^{1/3}} \qquad (11)$$

The variation of $l_{m\_critical}$ with temperature is plotted in Fig. 11. With the increase in temperature, $l_{m\_critical}$ increases. It is found that $l_{m\_critical}$ becomes negative when $T < T_{critical} = 387.58$ K, which explains why there is no critical $l_m$ at $T = 323.15$ K. Eq. (10) indicates that when $l_m \to 0$,

$k_{eff} \rightarrow \left[ 1/(l_m/l_p + l_p/l_m + 2) + 1/(l_m/l_p + 1) \right] k_p$. Through equating this limit to UO$_2$ thermal conductivity, the critical temperature under which no critical $l_m$ exists can be derived. As shown in Fig. 12, with the increase in BeO content, the critical temperature increases. To summarize, the condition for the enhancement of UO$_2$ thermal conductivity by doping with continuous BeO is

$$\begin{cases} l_m > l_{m\_critical}, & \text{when } T > T_{critical}, \\ \text{any } l_m, & \text{when } T \leq T_{critical}. \end{cases} \quad (12)$$

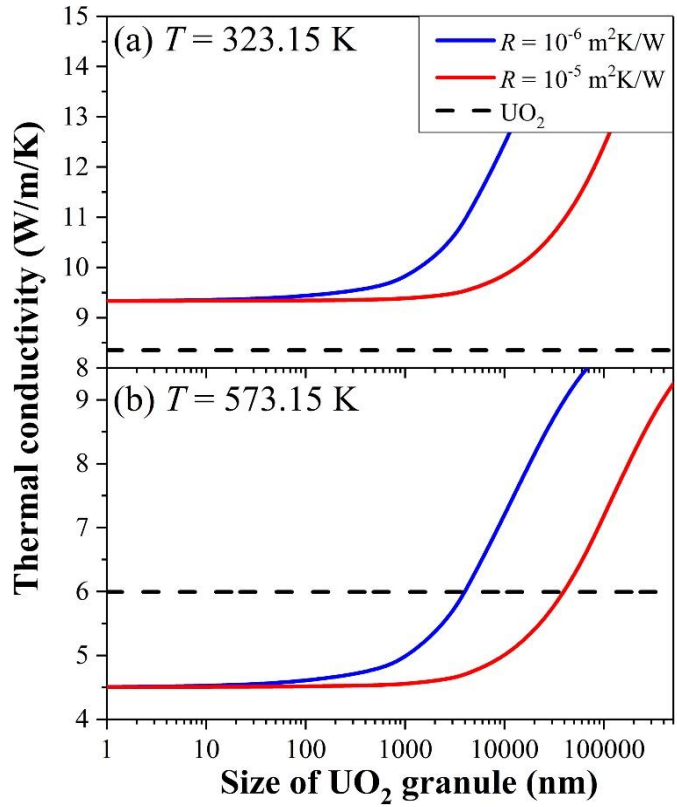

**Fig. 10.** Variation of the thermal conductivity of UO$_2$ containing 5 vol.% continuous BeO with respect to the size of UO$_2$ granule.

Furthermore, the conditions for the targeted improvement of UO$_2$ thermal conductivity by doping with continuous BeO are derived. If UO$_2$ thermal conductivity was to be increased by at least $\xi$, that is $k_{eff} \geq (1+\xi)k_m$, the following conditions should be satisfied according to Eq. (10):

$$(l_{mp}^2 - l_{mp} - 1) + \xi l_{mp}(1+l_{mp})^2 + k_m(l_{mp}+1)^2(1+\xi)/k_p - k_p(2l_{mp}^2 + l_{mp})/k_m \leq 0, \quad (13)$$

$$l_m \geq \frac{Rl_{mp}\left[-2k_m\left(1+l_{mp}\right)^2\left(1+\xi\right)+k_p\left(2+4l_{mp}\right)\right]}{\left(l_{mp}^2-l_{mp}-1\right)+\xi l_{mp}\left(1+l_{mp}\right)^2+k_m\left(l_{mp}+1\right)^2\left(1+\xi\right)/k_p-k_p\left(2l_{mp}^2+l_{mp}\right)/k_m} \quad .(14)$$

Eq. (13) expresses the condition that the volume fraction of BeO should be larger than a critical value. For example, if $UO_2$ thermal conductivity was to be improved by 80% at 1000 K, the volume fraction of BeO needed to be larger than 8.76%. If in the real experiments, 10 vol.% BeO was inserted, $UO_2$ granule size needed to be larger than 56.89 μm and 568.87 μm for achieving 80% enhancement of the thermal conductivity when $R = 10^{-6}$ and $10^{-5}$ $m^2K/W$, respectively.

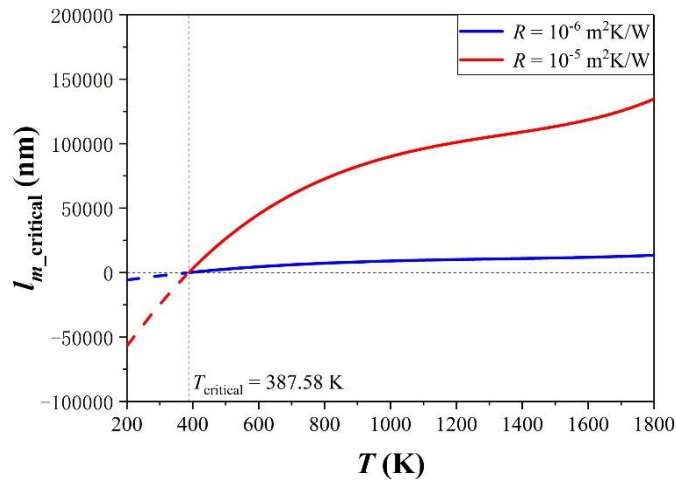

**Fig. 11.** Variation of the critical granule size of $UO_2$ with respect to the temperature for $R = 10^{-6}$ and $10^{-5}$ $m^2K/W$, respectively. The volume fraction of BeO is 5 vol.%.

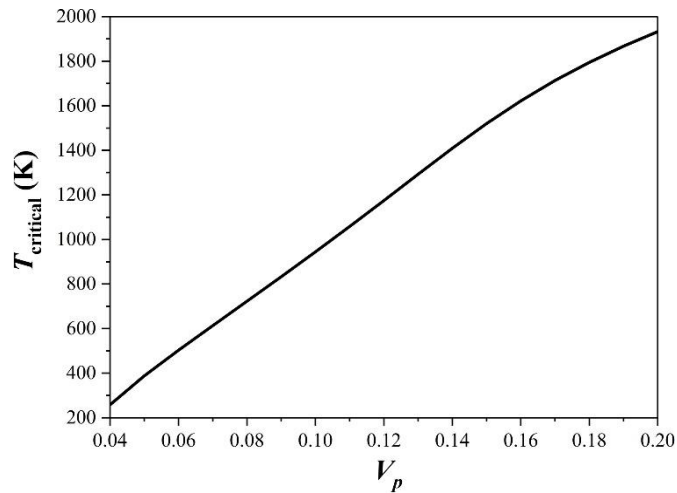

**Fig. 12.** Variation of the critical temperature with respect to the volume fraction of BeO. When the temperature is smaller than the critical temperature, the thermal conductivity of $UO_2$

containing continuous BeO is always larger than that of $UO_2$.

## 4. Conclusions

$UO_2$/BeO ITRs obtained through DMM based on different phonon dispersions are calculated and compared. Then, the impacts of ITR on $UO_2$-BeO thermal conductivity are investigated.

DMM based on Debye approximation and phonon dispersions predicted by DFT gives values of $UO_2$/BeO ITRs that are on the same order of magnitude at low temperatures ($T < 45$ K). However, at high temperatures ($T > 70$ K), predictions of DMM based on Debye approximation and acoustic modes ($10^{-8}$ $m^2K/W$) become one order of magnitude larger than that the predictions of DMM based on FB dispersions ($10^{-9}$ $m^2K/W$). This result can be explained through the density of states of $UO_2$ and BeO, which mainly overlay in the region of frequencies that are dominated by optic modes in $UO_2$ and acoustic modes in BeO. Therefore, at high temperatures, the activated phonons of optic modes in $UO_2$ become the main carriers of heat through $UO_2$/BeO interface, thus leading to the failure of DMM using only acoustic modes. These comparisons and analyses indicate that FB phonon dispersions are necessary for predicting $UO_2$/BeO ITR in the reactor, where the operating temperature is high.

To examine if DMM prediction is applicable to the real interface, comparison is made between $UO_2$-BeO thermal conductivities calculated by theoretical models that can consider ITR effect and experimental data. For $UO_2$ containing dispersed BeO, H-J model for a matrix with a random distribution of spherical dispersions is used. It is found that the results of H-J model using the ITR predicted by DMM can fit experimental data well, and the fitted BeO radius is on the same order of magnitude with that reported by the literature. This indicates that the vibrational mismatch between $UO_2$ and BeO considered by DMM is the major mechanism for attenuating the heat flux through $UO_2$/dispersed-$UO_2$ interface. Furthermore, it is found that the existence of ITR leads to the dependence of $UO_2$-BeO thermal conductivity on the size of dispersed BeO. With the decrease in BeO size, $UO_2$-BeO thermal conductivity decreases. There exists a critical value for BeO size. BeO dispersions with size smaller than this critical value cannot enhance the thermal conductivity of the fuel. The equation for this critical BeO size is derived. In addition, the conditions for the targeted enhancement of $UO_2$ thermal conductivity by doping with dispersed BeO are presented.

For $UO_2$ containing continuous BeO, a thermal conductivity model that can consider ITR effect is derived. It is found that using the ITR predicted by DMM, this model overestimates the thermal conductivities compared with the experimental data. If the calculated thermal conductivity is to be in agreement with experimental data, the ITR should be on the order of $10^{-6} \sim 10^{-5}$ m$^2$K/W. This result indicates that DMM is not applicable to the interface between $UO_2$ and continuous BeO. Therefore, the heat flux through $UO_2$/continuous-BeO interface is not mainly attenuated by vibrational mismatch between $UO_2$ and BeO, and there may exist voids on the interface that largely attenuates the flux. Furthermore, it is found that the presence of ITR results in the dependence of $UO_2$-BeO thermal conductivity on the size of $UO_2$ granule surrounded by continuous BeO. With the decrease in $UO_2$ granule size, $UO_2$-BeO thermal conductivity decreases. However, $UO_2$-BeO thermal conductivity is not necessarily smaller than $UO_2$ thermal conductivity when the size of $UO_2$ granule is small enough. It is found that under a critical temperature, $UO_2$-BeO thermal conductivity is always larger than $UO_2$ thermal conductivity. Above the critical temperature, $UO_2$-BeO thermal conductivity is larger than $UO_2$ thermal conductivity only when $UO_2$ granule size is larger than a critical value. The equation for this critical granule size is presented. And the conditions for the targeted enhancement of $UO_2$ thermal conductivity by inserting continuous BeO are derived.

To summarize, $UO_2$/BeO ITR predicted in this work can be used to predict the thermal conductivity of $UO_2$-BeO composite. And the conditions for the targeted enhancement of $UO_2$ thermal conductivity by doping with BeO proposed in this work can help to design the distribution, content, size of BeO and the size of $UO_2$ granule.

**Declaration of competing interest**

The authors declare no conflicts of interests.

**Acknowledgements**

This work was jointly supported by the National Key Research and Development Program of China (Grant No. 2017YFB0702401 and 2016YFB0201204) and the Science Challenge Project (Grant No. TZ2018002).

**Data Availability Statement**

The raw/processed data required to reproduce these findings cannot be shared at this time due to legal or ethical reasons.